\documentstyle[12pt]{article}
\begin{document}
\parskip 10pt plus 1pt
\title{Van der Waerden's Colouring Theorem and Weak-Strong Duality 
on the Lattice}
\author{
{\it  Debashis Gangopadhyay and Ranjan Chaudhury }
\\{S.N.Bose National Centre For Basic Sciences}\\
{JD Block, Sector-III, Salt Lake, Calcutta-700091, INDIA}\\
}
\baselineskip=20pt
\date{}
\maketitle

\begin{abstract}
Van der Waerden's (VDW) colouring theorem in combinatoric number 
theory [1] has scope for physical applications.The solution of the two colour 
case has  enabled the construction of an explicit mapping of an infinite, 
one dimensional antiferromagnetic Ising system to an effective pseudo-
ferromagnetic one with the coupling constants in the two cases becoming 
related [2].Here the three colour problem is solved and the results are used 
to obtain new insights in the theory of complex lattices, particularly those 
relating to ternary alloys. The existence of these mappings of a multicolour 
lattice onto a monochromatic one with different couplings illustrates  
{\it a new form of duality}.

PACS NO: 75.10 Hk
\end{abstract}
\vskip 2in
\newpage
One version of Van der Waerden's Colouring Theorem is [1] :

Let $X$ be a finite subset of $N^{d}$. For each m-colouring of $N^{d}$ there
exists a positive integer $\alpha$ and a point $\beta$ in $N^{d}$ such that 
the set $\alpha X + \beta$ is monochromatic. Moreover, the number $\alpha$ 
and the coordinates of the point $\beta$ are bounded by a function that 
depends only on $X$ and $m$ (and not on the particular colouring used).

Here we discuss the three colour case and show how a new type of weak-strong
duality on the lattice can be realised. We first briefly review the two colour
problem [2] . Next we give the solution to the three colour case and outline 
how this may be applicable to complex lattices and ternary alloys. We conclude 
by explaining the new type of weak-strong duality  evident in our approach
and list some more areas of application. Throughout we shall be using Ising-type
nearest-neighbour interactions.

Consider the antiferromagnetic spin ordering in an infinite one dimensional 
lattice with lattice spacing $b$.

We consider $N^{1}$ and our subset $X$ consists of
any two points of the one dimensional lattice such that the
first point has ``up'' spin (labelled by $R$ i.e. red) and the second
a ``down'' spin (labelled by $B$ i.e. blue).If $k$ denotes the point with 
an ``up'' spin then it necessarily follows that the ``down'' spin will occur
at a spacing $k + (2m+ 1)b, m = 0,1,2,..$. The set can be anywhere on the lattice.
The coupling between any two spins is taken to be some function of their
separation.To simulate the infinite lattice we use periodic boundary
conditions i.e. R..B..R..B..etc. Then the VDW theorem can be quantified as follows :
$$\left(\begin{array}{l}
R_{\alpha k+\beta}\\
R_{\alpha[k+(2m+1)b]+\beta}\\
\alpha k + \beta\\
\alpha[k+(2m+1)b]+\beta\end{array}\right) = 
\left(\begin{array}{cccc}
V_{11} & V_{12} & V_{13} & V_{14}\\
V_{21} & V_{22} & V_{23} & V_{24}\\
V_{31} & V_{32} & V_{33} & V_{34}\\
V_{41} & V_{42} & V_{43} & V_{44}\end{array}\right) 
\left(\begin{array}{l}
R_{k}\\
B_{k+(2m+1)b}\\
k\\
k+(2m+1)b \end{array}\right) \eqno(1)$$
For generality the lattice spacing in the subset $X$ is taken to be
$(2m + 1)b$\enskip   ($m = 0,1,2,...$).We are considering the case where
$(R..B..R..B...) \rightarrow (R..R..R..R..)$.

We have combined the colour labels (``up'' and ``down'') and the
position labels ($k$ and $k+(2m+1)b$) into a single column
vector. The column vector on the right hand side means that
there is an ``up'' spin $(R)$ at  position $k$ and a ``down''
spin $(B)$ at  position $k+(2m+1)b$ where $(2m+1)b$ is the lattice separation.
The column vector on the left hand side has the corresponding meaning. 
The ordering we have used in the initial configuration is (R, B). 
The solution to the $V$ matrix is
$$V = \left(\begin{array}{cccc}
1 & 0 & 0 & 0\\
1 & 0 & 0 & 0\\
0 & 0 & \alpha - \beta/(2m+1)b & \beta/(2m+1)b\\
0 & 0 & - \beta/(2m+1)b & \alpha + \beta/(2m+1)b
\end{array}\right) \eqno(2)$$
Now consider the case $R \Leftrightarrow B$ in equation $(1)$. The
matrix $V$ remains the same. This is expected as this reflects a
symmetry of the system. The Van der Waerden transformation
should be insensitive to what is called ``up'' and what is
called ``down''.The final configuration can also consist of all ``down'' 
spins starting with the initial ordering as (R, B). Then also the VDW matrix
can be determined [2].

As the final configuration is monochromatic, it can be easily
verified that for the system we are considering with periodic
boundary conditions, $\alpha = 2$ and $\beta = (2m + 1)b$. In
fact, it can be shown that the latter part of the theorem can
also be realised viz. $\alpha$ and $\beta$ are bounded by a
function  that depends only on $X$ and $m$.

The energy of the initial configuration $E_{i}$
may be written formally as some average of some hamiltonian $H =
\sum_{ij(i \not= j)} s_{i} J_{ij} s_{j}$ with nearest neighbour interactions
$$E_{i} = \displaystyle\sum_{{m,n}_{m \not= n}} < m \vert H \vert
n > \eqno(3)$$
The suggestive notation used in $(3)$ is for convenience. To
evaluate this split the chain into blocks and expand $(3)$ in terms of these 
blocks. The block energies are all identical. If $N$ be the number of blocks 
$(3)$ becomes :
$$\frac{E_{i}}{N} = <u^{T}_{1} \vert H \vert u_{2} >=
J_{21}((2m+1)b) B_{k+(2m+1)b} R_{k+2(2m+1)b} \eqno(4a)$$
where
$$\vert u_{1} > = 
\left(\begin{array}{l}
R_{k}\\
B_{k+(2m+1)b}\\
k\\
k+(2m+1)b \end{array}\right),
\vert u_{2} > = 
\left(\begin{array}{l}
R_{k+2(2m+1)b}\\
B_{k+3(2m+1)b}\\
k + 2(2m+1)b\\
k + 3(2m+1)b
\end{array}\right) \eqno(4b)$$
$$H = \left(\begin{array}{cccc}
0 & 0 & 0 & 0\\
J_{21} & 0 & 0 & 0\\
0 & 0 & 0 & 0\\
0 & 0 & 0 & 0
\end{array}\right) \eqno(4c)$$
The subscripts in $(4a)$ are only there to show that these are
nearest neighbour interactions.

In a similar manner it may be shown that the final energy is 
$$\frac{E_{f}}{N'} = <v^{T}_{1}\vert H'\vert v_{2}>
\phantom{.....................}$$
$$\phantom{\frac{E_{f}}{N'}...............................} = J'_{21} (\alpha(2m + 1)b)
R_{\alpha[k+(2m+1)b]+\beta} R_{\alpha[k+2(2m+1)b]+\beta} \eqno(5a)$$
$$\vert v_{1} >= \left(\begin{array}{l}
R_{\alpha k+\beta}\\
R_{\alpha[k+(2m+1)b]+\beta}\\
\alpha k + \beta\\
\alpha[k+(2m+1)b]+\beta\end{array}\right),
\vert v_{2} >= 
\left(\begin{array}{l}
R_{\alpha[k+2(2m+1)b]+\beta}\\
R_{\alpha[k+3(2m+1)b]+\beta}\\
\alpha[k + 2(2m+1)b]+\beta\\
\alpha[k + 3(2m+1)b]+\beta
\end{array}\right) \eqno(5b)$$
$$H' = \left(\begin{array}{cccc}
0 & 0 & 0 & 0\\
J'_{21} & 0 & 0 & 0\\
0 & 0 & 0 & 0\\
0 & 0 & 0 & 0
\end{array}\right) \eqno(5c)$$
The antiferromagnetic chain has been mapped onto itself except that now 
there is an effective ferromagnetic ordering. No dynamics is  involved
and so the initial and final energies are equal. As usual, in the calculation 
of energies we take the value of the ``up'' spin to be $+1$ and that for the
``down'' spin to be $-1$. The lattice is infinite, so $N = N'$. Equating
$(4a)$ with $(5a)$ we get :
$$J'_{21}(\alpha(2m + 1)b) = - J_{21}((2m + 1)b)$$
i.e.
$$J'_{21}(2(2m + 1)b) = - J_{21}((2m + 1)b) \eqno(6)$$
Hence the crucial result of [2] was that {\it the antiferromagnetic system is
in some sense dual to a pseudo-ferromagnetic one with a coupling opposite in 
sign and a function of $(\alpha)^{-1}$ times the lattice unit of the 
pseudo-ferromagnetic one}. The nature of this duality will become more clear
in the solution to the three colour case which now follows.

Consider $N^{1}$ with the subset $X$ consisting of any three points of the 
one dimensional lattice such that the first point has  colour "red" ($R$), 
the second point has colour "blue" ($B$) and the third has colour
"green" ($G$).If $k$ denotes the point with a "red" colour then it necessarily 
follows that the subsequent "blue" and "green" colours will occur at a spacing 
$k + (3m - 2)b$ and $k + 2(3m -2)b$ respectively, $ m = 1,2,..$.The set can 
be anywhere on the lattice.The coupling between any two colours is taken to be
some function of their separation.As before, we use periodic boundary conditions 
i.e. R..B..G..R..B..G..etc. Then the VDW theorem can be quantified as:
$$\left(\begin{array}{l}
R_{\alpha k+\beta}\\
R_{\alpha[k+(3m-2)b]+\beta}\\
R_{\alpha[k+2(3m-2)b]+\beta}\\
\alpha k + \beta\\
\alpha[k+(3m-2)b]+\beta\\
\alpha[k+2(3m-2)b]+\beta\end{array}\right) =$$ 
$$\left(\begin{array}{cccccc}
V_{11} & V_{12} & V_{13} & V_{14} & V_{15} & V_{16}\\
V_{21} & V_{22} & V_{23} & V_{24} & V_{25} & V_{26}\\
V_{31} & V_{32} & V_{33} & V_{34} & V_{35} & V_{36}\\
V_{41} & V_{42} & V_{43} & V_{44} & V_{45} & V_{46}\\    
V_{51} & V_{52} & V_{53} & V_{54} & V_{55} & V_{56}\\
V_{61} & V_{62} & V_{63} & V_{64} & V_{65} & V_{66}\end{array}\right)
\left(\begin{array}{l}
R_{k}\\
B_{k+(3m-2)b}\\
G_{k+2(3m-2)b}\\
k\\
k+(3m-2)b\\
k+2(3m-2)b\end{array}\right)\eqno(7)$$
We are considering the case $(R..B..G..R..B..G..) \rightarrow (R..R..R..R..)$.

Note that for $\beta = 0$, the VDW matrix in the two colour case is an 
orthogonal matrix. Consistency of the formalism demands that this property 
should also be valid for the VDW matrix in the three colour case. Using this
property, some straightforward algebra yields two solutions for the VDW matrix 
on right hand side of $(7)$. These are :
\newpage
$$ V_{1} = \left(\begin{array}{cccccc}
1 & 0 & 0 & 0 & 0 & 0\\
1 & 0 & 0 & 0 & 0 & 0\\
1 & 0 & 0 & 0 & 0 & 0\\
0 & 0 & 0 & \alpha -\beta/2(3m-2)b & 0 & \beta/2(3m-2)b\\    
0 & 0 & 0 & -\beta/2(3m-2)b &\alpha & \beta/2(3m-2)b\\
0 & 0 & 0 & -\beta/2(3m-2)b & 0 & \alpha+\beta/2(3m-2)b \end{array}\right)\eqno(8a)$$
and\\
$ V_{2} =$\\
$$\left(\begin{array}{cccccc}
1 & 0 & 0 & 0 & 0 & 0\\
1 & 0 & 0 & 0 & 0 & 0\\
1 & 0 & 0 & 0 & 0 & 0\\
0 & 0 & 0 &(2/3)\alpha-\beta/2(3m-2)b & (2/3)\alpha &-(1/3)\alpha+\beta/2(3m-2)b\\    
0 & 0 & 0 &(2/3)\alpha-\beta/2(3m-2)b &-(1/3)\alpha &(2/3)\alpha+\beta/2(3m-2)b\\
0 & 0 & 0 &-(1/3)\alpha-\beta/2(3m-2)b 
&(2/3)\alpha &(2/3)\alpha+\beta/2(3m-2)b\end{array}\right)$$
$$\eqno(8b)$$
As we have periodic boundary conditions, the last lattice
site at infinity is identified with the starting (first) site.So
the initial energy $E_{i}$ of the system can be written as :
$$E_{i}
=(energy \hskip.1in of\hskip.1in primitive\hskip.1in 3-cell)
\hskip.1in\times\hskip.1in N$$
$$=N[R_{k}J_{RB}((3m-2)b))B_{k+(3m-2)b}$$ 
$$+B_{k+(3m-2)b}J_{BG}((3m-2)b))G_{k+2(3m-2)b}]$$
$$+G_{k+2(3m-2)b}J_{GR}((3m-2)b))\hskip.1inR_{k+3(3m-2)b}$$
i.e.
$$E_{i}/N = J_{RB}((3m-2)b)\hskip.1inR_{k}\hskip.1inB_{k+(3m-2)b}$$ 
$$ + J_{BG}((3m-2)b)\hskip.1inB_{k+(3m-2)b}\hskip.1inG_{k+2(3m-2)b}$$
$$ + J_{GR}((3m-2)b)\hskip.1inG_{k+2(3m-2)b}\hskip.1inR_{k+3(3m-1)b}\eqno(9a)$$

Equation $(9a)$  can be written as
$$E_{i} = \displaystyle\sum_{{m,n}_{m \not= n}} < m \vert H \vert
n > \eqno(9b)$$
To evaluate this split the chain (as before)into blocks and expand $(9b)$ in 
terms of these blocks. The block energies are all identical. If $N$ be the 
number of blocks $(9b)$ becomes :
$$E_{i}/N = <u^{T}_{1} \vert H \vert u_{2}>=J_{RB}((3m-2)b)\hskip.1inR_{k}\hskip.1inB_{k+(3m-2)b}$$ 
$$ + J_{BG}((3m-2)b)\hskip.1inB_{k+(3m-2)b}\hskip.1inG_{k+2(3m-2)b}$$
$$ + J_{GR}((3m-2)b)\hskip.1inG_{k+2(3m-2)b}\hskip.1inR_{k+3(3m-2)b}\eqno(9c)$$
where
$$\vert u_{1} > = 
\left(\begin{array}{l}
R_{k}\\
B_{k+(3m-2)b}\\
G_{k+2(3m-2)b}\\
k\\
k+(3m-2)b\\
k+2(3m-2)b\\ \end{array}\right),
\vert u_{2} > = 
\left(\begin{array}{l}
R_{k+3(3m-2)b}\\
B_{k+4(3m-2)b}\\
G_{k+5(3m-2)b}\\
k+3(3m-2)b\\
k+4(3m-2)b\\
k+5(3m-2)b\\
\end{array}\right) \eqno(10a)$$
$$H = (1/2)\left(\begin{array}{cccccc}
0 & J_{BR} & J_{GR} & 0 & 0 & 0\\
J_{RB} & 0 & J_{GB} & 0 & 0 & 0\\
J_{RG} & J_{BG} & 0 & 0 & 0 & 0\\
0 & 0 & 0 & 0 & 0 & 0\\
0 & 0 & 0 & 0 & 0 & 0\\
0 & 0 & 0 & 0 & 0 & 0\\
\end{array}\right) \eqno(10b)$$
Here we have assumed that $J_{RB}=J_{BR},J_{RG}=J_{GR},J_{BG}=J_{GB}$. 

Similarly
$$E_{f}/N'=<v^{T}_{1} \vert H' \vert v_{2}>
=J_{RR}'(\alpha(3m-2)b)\hskip.1inR_{\alpha(k+(3m-2)b)+\beta}\hskip.1inR_{\alpha(k+2(3m-2)b)+\beta}\eqno(11a)$$
where
$$\vert v_{1} > = 
\left(\begin{array}{l}
R_{\alpha k+\beta}\\
R_{\alpha(k+(3m-2)b)+\beta}\\
R_{\alpha(k+2(3m-2)b+\beta}\\
\alpha k+ \beta\\
\alpha(k+(3m-2)b)+\beta\\
\alpha(k+2(3m-2)b)+\beta\\ \end{array}\right),
\vert v_{2} > = 
\left(\begin{array}{l}
R_{\alpha(k+3(3m-2)b)+\beta}\\
R_{\alpha(k+4(3m-2)b)+\beta}\\
R_{\alpha(k+5(3m-2)b)+\beta}\\
\alpha(k+3(3m-2)b)+\beta\\
\alpha(k+4(3m-2)b)+\beta\\
\alpha(k+5(3m-2)b)+\beta\\
\end{array}\right) \eqno(11b)$$
$$H' = (1/2)\left(\begin{array}{cccccc}
0 & J_{RR}' & J_{RR}' & 0 & 0 & 0\\
J_{RR}' & 0 & J_{RR}' & 0 & 0 & 0\\
J_{RR}' & J_{RR}' & 0 & 0 & 0 & 0\\
0 & 0 & 0 & 0 & 0 & 0\\
0 & 0 & 0 & 0 & 0 & 0\\
0 & 0 & 0 & 0 & 0 & 0\\
\end{array}\right) \eqno(11c)$$
In order to calculate the energies quantitatively,let us assign the values 
$r,c$ and $g$ to the colours $R,B$ and $G$ respectively. These are like 
{\it weight factors} that distinguish the various colours from each other.
(They are analogues of the values $+1$ and $-1$ in the case of "up" spin and 
"down" spin respectively.)

Then equating the initial and final energies and noting that for the infinite
lattice $N=N'$ (as before) we have 
$$r^2\hskip.05inJ_{RR}'(\alpha(3m-2)b)=rc\hskip.01inJ_{RB}((3m-2)b)$$
$$+ cg\hskip.01in J_{BG}((3m-2)b)$$
$$+ gr\hskip.01inJ_{GR}((3m-2)b)\eqno(12a)$$
Two other mappings (to a monochromatic set) are also possible, {\it viz.},
$B..B..B..$ and $G..G..G..$.Therefore :
$$c^2\hskip.025inJ_{BB}'(\alpha(3m-2)b)=rc\hskip.025inJ_{RB}((3m-2)b)$$
$$+cg\hskip.025in J_{BG}((3m-2)b)+gr\hskip.025inJ_{GR}((3m-2)b)\eqno(12b)$$
$$g^2\hskip.05inJ_{GG}'(\alpha(3m-2)b)=rc\hskip.025inJ_{RB}((3m-2)b)$$
$$+ cg\hskip.025in J_{BG}((3m-2)b) + gr\hskip.025inJ_{GR}((3m-2)b)\eqno(12c)$$
These equations imply
$$r^2/c^2=J_{BB}'/J_{RR}'\hskip.025in,\hskip.1inr^2/g^2\\
=J_{GG}'/J_{RR}'\hskip.025in,\hskip.1inc^2/g^2=J_{GG}'/J_{BB}'\eqno(13)$$
We have thus related the couplings in the two configurations.Note that it 
is quite plausible that the weights $r,c$ and $g$ can be empirically 
determined.If just one of the the self-couplings $J_{RR}', J_{BB}'$ and $J_{GG}'$ 
can be estimated then the others can be determined from eq.$(13)$. The fact
that the weight factors occur quadratically in $(12)$ and $(13)$ allows a
freedom in the sign of the couplings and so a variety of physical systems 
becomes amenable to treatment via our formalism. Moreover, (under certain 
approximations) using the two colour solution $J_{RB}$ can be related to 
$J_{RR}'$ and $J_{BB}'$; $J_{RG}$ can be related to $J_{RR}'$ and $J_{GG}'$ 
and $J_{BG}$ can be related to $J_{BB}'$ and $J_{GG}'$ and so on.  What 
we are trying to stress is that the set of equations $(12)$ can, in principle, 
be solved by a judicious usage of the two-colour and three-colour solutions 
and some empirical or experimental inputs.Thus the entire theory of disorders
can be studied in a new light. Moreover, the VDW matrices $V$ have interesting
properties which can be used to study various other aspects of complex lattices.
These will be reported elsewhere.

The results of this letter bring to light a new form of duality on the lattice.
A system with multiple degrees of freedom has been mapped onto a system
with only a single effective degree of freedom and this is possible when the 
couplings are related in some particular way. Note that the lattice spacing 
dependence of the monocolour couplings is $\alpha$ times that of the 
multicolour couplings. Now, $\alpha$ is always an integer (here it works out 
to be $3$). {\it Hence a possibility exists , depending on the exact values of the 
weight factors, of the monocolour coupling becoming strong (or weak) relative 
to some or all of the multicolour ones. This, therefore, is an example of 
weak-strong duality in an otherwise purely classical system.}

Finally, we list some of the other possible applications of the results of 
this letter :

(i)In realistic situations such as in ternary alloys or in ionic (covalent)
lattice in one dimension containing three ions (atoms) per unit cell the 
weight factors ($r,c,g$) may be identified with the effective charge on the
ion or the number of valence electrons on the atom.$J$ is the separation
dependent bare interaction potential between near-neighbour dissimilar ions
(atoms) and $J'$ is the effective interaction potential between near-neighbour
similar ions (atoms).Eqs.$(12)$ and $(13)$ show that a suitable combination of
cations and anions can even induce an {\it effective attraction} between cations
(anions) themselves.

(ii)Our approach based on the VDW mapping has a formal similarity to "Madelung's
method" [3] of determining the cohesive energy of an ionic crystal.Besides, this 
approach may also be looked upon as an analogue of "dielectric function" [4]
formalism at the classical level.The effective interaction calculated may be
regarded as a "pseudo-potential".

(iii)The $3-$ colour  solutions have possible applications
to $c-$axis properties of the layered high-$T_{C}$ oxide systems; $1$-d or 
quasi-$1$-d organic compounds and $1$-d spin glass with competing interaction
and frustration.

(iv)Our approach may be generalised and slightly modified to deal with phase
separation problems.

(v)The method developed here is also useful in certain approaches to quantum 
gravity.

\end{document}